\documentclass[aps,prd,twocolumn,showpacs,floatfix,preprintnumbers,amsmath,amssymb,nofootinbib,superscriptaddress]{revtex4-1}

\usepackage{graphics}
\usepackage{graphicx}
\usepackage{colordvi}
\usepackage{color}
\usepackage{mathtools}
\usepackage{empheq}
\usepackage{bm}%
\usepackage{multirow}
\usepackage{hyperref}

\def\be{\begin{equation}}
\def\ee{\end{equation}}
\def\ba{\begin{eqnarray}}
\def\ea{\end{eqnarray}}
\def\dd{\textrm{d}}

\begin{document}

\title{Probing primordial features with the primary CMB}

\author{Mario Ballardini}\email{mario.ballardini@gmail.com}
\affiliation{Department of Physics and Astronomy, University of the Western Cape, 
Cape Town 7535, South Africa}
\affiliation{INAF/OAS Bologna, via Gobetti 101, I-40129 Bologna, Italy}

\date{\today}

\begin{abstract}
%
We propose to study the imprint of features in the primordial power spectrum with 
the primary CMB after the subtraction of the reconstructed ISW signal from the 
observed CMB temperature angular power spectrum. 
We consider the application to features models able to fit two of the large scales 
anomalies observed in the CMB temperature angular power spectrum: the deficit of power 
at $\ell \sim 2$ and at $\ell \sim 22$. 

We show that if the features comes from the primordial power spectrum we should be 
find consistent constraints of these features model from the CMB temperature angular 
power spectrum removing or not the late ISW signal.
Moreover, this method shows also some improvement on the constraints on the features parameters 
up to $16\%$ for models predicting a suppression of power of the quadrupole and up to 
$27\%$ for models with features at $\ell \sim 22$, assuming instrumental sensitivity 
similar to the $Planck$ satellite (depending on the goodness of the ISW reconstruction).
Furthermore, it gives the opportunity to understand if these anomalies are attributed 
to early- or late-time physics.
\end{abstract}

\maketitle

\section{Introduction}
Although observations show how a spatially flat $\Lambda$CDM model with a tilted 
power-law spectrum of primordial density fluctuations provides a good fit to CMB temperature 
and polarization anisotropies \cite{Akrami:2018vks}, there are interesting hints for new physics 
beyond the $\Lambda$CDM model based on slow-roll inflation in the WMAP \cite{Peiris:2003ff} 
and $Planck$ data \cite{Planck:2013jfk,Ade:2015lrj,Akrami:2018odb}, such as anomalies in the large angular 
scale pattern of CMB temperature anisotropies \cite{Peiris:2003ff,Covi:2006ci,Planck:2013jfk,Benetti:2013cja,Miranda:2013wxa,Easther:2013kla,Chen:2014joa,Achucarro:2014msa,Hazra:2014goa,Hazra:2014jwa,Hu:2014hra,Ade:2015lrj,Gruppuso:2015zia,Gruppuso:2015xqa,Hazra:2016fkm,Torrado:2016sls,Obied:2018qdr,Akrami:2018odb}.

Anomalies in the CMB angular power spectra, as well as in the dark matter power spectrum, 
are predicted by several theoretically well motivated mechanisms that occur during inflation; 
these mechanisms support deviations from a simple power-law for the primordial power spectrum, 
connected with the violation of the slow-roll phase, and provide a better fit to the CMB data 
at $\sim 2\sigma$.

\begin{figure}
\centering
\includegraphics[width=.5\textwidth]{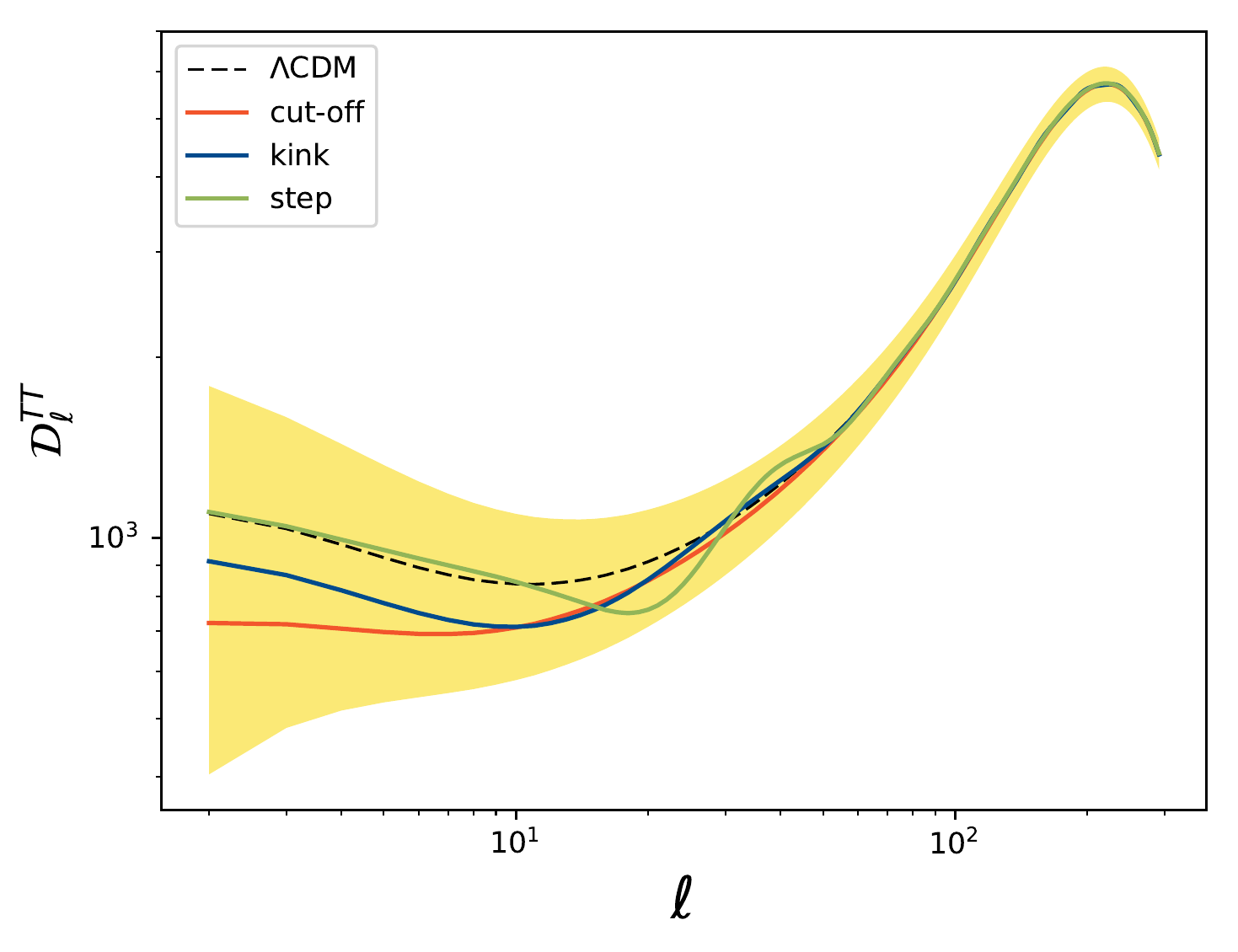}
\caption{CMB temperature angular power spectrum best-fit for $\Lambda$CDM (dashed black), 
{\bf cut-off} (red), {\bf kink} (blue), {\bf step} (green) models from 
$Planck$ 2015 TT+lowP data \cite{Ade:2015lrj}. 
The yellow band represents the error bar from the cosmic-variance only for $\Lambda$CDM.}
\label{fig:variance}
\end{figure}

In Fig.~\ref{fig:variance}, it is plotted the comparison between the best-fit CMB temperature 
power spectrum for the standard $\Lambda$CDM model and the best-fits for some features models 
\cite{Ade:2015lrj} which improve the fit of CMB data. Although the difference between these 
models, the cosmic-variance restricts our ability to discriminate between them even with 
a perfect measure of the CMB anisotropies.

The situation improves if well-suited data in addition to the CMB temperature anisotropies 
are available:
\begin{itemize}
\item CMB $E$-mode polarization have been highlighted as a possible way to 
constrain  primordial features with high confidence thanks to the narrower trasfer functions 
compared to the ones of the CMB temperature \cite{Mortonson:2009qv,Chluba:2015bqa,Finelli:2016cyd,Hazra:2017joc}.
\item The opportunity to look elsewhere for the imprint of primordial features, as in the 
matter power spectrum, is a unique chance to improve our current understanding of these 
possible anomalies; see for instance 
\cite{Zhan:2005rz,Huang:2012mr,Chen:2016vvw,Chen:2016zuu,Ballardini:2016hpi,Xu:2016kwz,Fard:2017oex,Palma:2017wxu,LHuillier:2017lgm,Ballardini:2017qwq}.
\item Combined search for primordial features in the power spectrum and bispectrum is 
another promising way to test such models thanks to the imprints on higher-order correlators 
\cite{Fergusson:2014hya,Fergusson:2014tza,Ade:2015ava,Meerburg:2015owa,Xu:2016kwz,MoradinezhadDizgah:2018ssw,Karagiannis:2018jdt}.
\end{itemize}

In this paper, we propose a further method to improve the current understanding of the 
large scales CMB anomalies based on the possibility to subtract the reconstructed 
integrated Sachs-Wolfe (ISW) signal from the observed CMB temperature angular power spectrum 
in order to constrain models with features in the primordial power spectrum with the primary CMB.
After subtracting the ISW signal, possible by cross-correlating CMB maps with tracer maps of 
the matter density fluctuation, we have the opportunity to test the CMB angular power spectrum 
dominated by the SW contribution at the largest scales like the primary CMB signal generated at 
the last scattering surface.
This technique has been already proposed and applied to real data to study the significance of 
anomalies at CMB maps level with and without the contamination from the late ISW signal 
\cite{Efstathiou:2009di,Francis:2009pt,Muir:2016veb,Copi:2016hhq}.

ISW, such as CMB lensing deflection, can be considered as foreground contribution to the 
primary CMB signal. They can be used to further study the information content from some 
late-time physics (dark energy and small scales matter perturbation for instance), but they 
also ceil part of the primary CMB signal introducing some degeneracies between primordial 
paramters and other ones.

The ISW contribution to the CMB temperature fluctuations in direction ${\bf \hat{n}}$ is 
a secondary anisotropy in the CMB caused by the passage of CMB photons through evolving 
gravitational potential wells
\be
\frac{\delta T_\mathrm{ISW}}{T} ({\bf \hat{n}}) = - \int \dd z\, e^{-\tau (z)} \left[  \frac{\dd \Phi}{\dd z} ({\bf \hat{n}}\,, z) + \frac{\dd \Psi}{\dd z} ({\bf \hat{n}}\,, z)  \right] \,,
\ee
where $\Phi$ and $\Psi$ are the gravitational potentials in the longitudinal gauge and 
$e^{-\tau (z)}$ is the visibility function. On large scales, in a dark-energy-dominated universe, 
CMB photons gain energy when they pass through the decaying potential wells associated with 
overdensities and lose energy on passing through underdensities \cite{Kofman:1985fp}. 
This effect mainly contributes to large angular scales and therefore at low multipoles, 
i.e. $\lesssim 100$, since there is a little power in the potentials at late times on 
scales that entered the Hubble radius during radiation domination.

\section{Primary CMB anisotropies temperature angular power spectrum}
The different components that source the observed CMB temperature are assumed to have a 
negligible correlation with the others and drawn from a Gaussian distributions with mean 
zero and a known covariance matrix. 
Therefore, the combination $T^{\rm obs}-T^{\rm ISW} = T^{\rm primary} + T^{\cal N} + T^{\rm fg}$ 
will be distributed as a Gaussian with covariance matrix equals to the sum of the covariance
matrices of the noise and the primordial CMB.

By knowing the ISW contribution to the CMB temperature anisotropies it is possible 
to reconstruct the primary CMB temperature angular power spectrum at large scales
\begin{align}
&C_\ell^{\rm primary} = C_\ell^{\rm TT} - C_\ell^{\rm ISW} \,,\\
\label{eqn:NoisePrimary}
&{\cal N}_\ell^{\rm primary} = {\cal N}_\ell^{\rm T} + {\cal N}_\ell^{\rm ISW} \,,
\end{align}
where ${\cal N}_\ell^{\rm ISW}$ is the noise of the ISW angular power spectrum after 
the reconstruction.

ISW is generally reconstructed by cross-correlating CMB temperature angular power spectrum 
with LSS galaxy surveys \cite{Crittenden:1995ak} or other LSS tracers such as CMB lensing 
\cite{Manzotti:2014kta}, termal Sunyaev-Zeldovich \cite{Taburet:2010hb}, intensity mapping 
emission lines \cite{Pourtsidou:2016dzn} and clusters of galaxies \cite{Ballardini:2017xnt}.

In order to quantifity the errors from the reconstruction of the ISW signal by 
cross-correlating the CMB with one or more LSS tracers, we consider the standard theoretical 
signal-to-noise ratio (defined according to \cite{Cooray:2001ab,Afshordi:2004kz}) to build 
the noise angular power spectra for three different cases: a 3$\sigma$ level reconstruction 
of the ISW signal, compatible with the significance obtained in \cite{Ade:2015dva} by 
cross-correlating the $Planck$ temperature map with a compilation of publicly available galaxy 
surveys \cite{Ade:2015dva}; a 6$\sigma$ significance expected for next-generation of LSS 
galaxy surveys \cite{Raccanelli:2015lca,Pourtsidou:2016dzn}; an ideal case with a perfect 
reconstruction ($\sim 10\sigma$) of the late-time ISW signal with ${\cal N}^{\rm ISW} \simeq 0$ 
in Eq.~\eqref{eqn:NoisePrimary}.

\section{Fisher forecast formalism}
With these definitions in hand, we can  proceed to perform a Fisher matrix analysis 
for CMB angular power spectra (temperature and $E$-mode polarization)~\cite{Knox:1995dq,Jungman:1995bz,Seljak:1996ti,Zaldarriaga:1996xe,Kamionkowski:1996ks}
\be
{\cal F}_{\alpha\beta}^\mathrm{CMB} = \frac{1}{2} \text{tr} \left[\mathbf{C}_{,\alpha} \mathbf{C}^{-1} \mathbf{C}_{,\beta} \mathbf{C}^{-1}\right] \,,
\ee
where
\be
\mathbf{C} =
\begin{bmatrix}
\bar{C}_\ell^{TT} & \bar{C}_\ell^{TE} \\
\bar{C}_\ell^{TE} & \bar{C}_\ell^{EE} \\
\end{bmatrix} \,.
\ee
Here $\bar{C}_\ell^{X}$ is the sum of the theoretical spectrum $C_\ell^{X}$ and the effective 
noise ${\cal N}^{X}_{\ell}$, which is given by the inverse noise weighted combination of the 
instrumental noise de-convolved with the beams of different frequency channels.
For the temperature and polarization angular power spectra, a noise power spectrum with Gaussian 
beam profile \cite{Knox:1995dq} has been used
\be
{\cal N}^{\rm X}_\ell=\sigma_{\rm X}\, b^{-2}_\ell \,.
\ee
Here $b_\ell^2$ is the beam window function, assumed Gaussian, with 
$b_\ell = e^{-\ell(\ell+1)\theta_{\rm FWHM}^2/16 \ln 2}$; $\theta_{\rm FWHM}$ is the full 
width half maximum (FWHM) of the beam in radians; $\sigma_{\rm T}$ and $\sigma_{\rm E}$ are 
the square of the detector noise level on a steradian patch for temperature and polarization, 
respectively.

\section{Models of features in the primordial power spectrum}
We consider three inflation models that generate features in the primordial power spectrum 
(see Fig.~\ref{fig:variance}): 
the {\bf cut-off} model \cite{Contaldi:2003zv} which reproduces a suppression of power at 
large scales, and two models which lead to localized features in the primordial power spectrum, 
i.e. the {\bf kink} model \cite{Starobinsky:1992ts} and the {\bf step} model \cite{Dvorkin:2009ne}. 
Following~\cite{Ballardini:2016hpi}, the fiducial spectra are centred at their best-fit 
parameters from $Planck$ 2015 TT+lowP data \cite{Ade:2015lrj} for each parameterization. 
The primordial power spectrum can be written as the standard power-law 
$\mathcal{P}_{\mathcal{R},0}$, modulated by the contribution dues to the violation of slow-roll
\begin{align}
\label{eqn:generalPPS}
&\mathcal{P}_{\mathcal{R}} (k) = \mathcal{P}_{\mathcal{R},0} (k) \cdot \mathcal{P}_{\mathcal{R},\,X}(k) \,,\\
&\mathcal{P}_{\mathcal{R},0} (k) = A_{\rm s}\left({k\over k_*}\right)^{n_{\rm s}-1} \,,
\end{align}
where $A_{\rm s}$ is the amplitude of the curvature power spectrum, $n_{\rm s}$ is the scalar 
spectral index and the pivot scale is fixed at $k_*=0.05$ Mpc$^{-1}$.

The non-canonical contribution to $\mathcal{P}_{\mathcal{R}}$ for the {\bf cutoff} model is given by
\begin{align}
\label{eqn:cutoff}
\mathcal{P}_{\mathcal{R},\,{\rm cutoff}}(y) &= 1 - e^{-y^{\lambda_{\rm c}}} \,,\\
y &\equiv {k\over k_{\rm c}}\,, 
\end{align}
for the {\bf kink} model by
\begin{align}
\mathcal{P}_{\mathcal{R},\,{\rm kink}}(y) &= 1 + \frac{9}{2}\mathcal{A}_\mathrm{kink}^2\left( \frac{1}{y} + \frac{1}{y^3} \right)^2 \notag\\
&+ \frac{3}{2}\mathcal{A}_\mathrm{kink}\left( 4 + 3\mathcal{A}_\mathrm{kink} - 3\frac{\mathcal{A}_\mathrm{kink}}{y^4} \right)^2 \frac{1}{y^2} \cos(2y) \notag\\
&+ 3\mathcal{A}_\mathrm{kink}\left( 1 - \frac{1 + 3\mathcal{A}_\mathrm{kink}}{y^2} - \frac{3\mathcal{A}_\mathrm{kink}}{y^4} \right)^2 \frac{1}{y} \sin(2y) \,,\\
y &\equiv {k\over k_\mathrm{kink}} \,,
\label{eqn:kink}
\end{align}
and for the {\bf step} model by
\begin{align}
\mathcal{P}_{\mathcal{R},\,{\rm step}}(y) &= \exp \Big\{ \mathcal{I}_0(y) + \ln \Big[1+\mathcal{I}_1^2(y)\Big] \Big\} \,,\\
y &\equiv {k\over k_\mathrm{step}}\,,
\end{align}
where the first- and second-order parts are
\ba
\mathcal{I}_0 (y) &=& {\cal A}_{\rm step} W'(y) \mathcal{D}\left(\frac{y}{ x_{\rm step}}\right)\,,
\label{eqn:step_first_order}\\
\sqrt{2}\,\mathcal{I}_1 (y) &=& \frac{\pi}{2}\left(1-n_{\rm s}\right) 
+ {\cal A}_{\rm step} X'(y) \mathcal{D}\left(\frac{y}{x_{\rm step}}\right)\,,
\label{eqn:step_second_order}
\ea
where a prime denotes ${\rm d}/{\rm d} \ln y$, and the damping envelope is
\be
\label{eqn:damping}
\mathcal{D}(y) = \frac{y}{\sinh y} \,.
\ee
The window functions are
\begin{align}
W(y) &= \frac{3\sin(2y)}{2y^3} - \frac{3\cos(2y)}{y^2} - \frac{3\sin(2y)}{2y} \,,\\
X(y) &= \frac{3}{y^3} \left(\sin y - y\cos y\right)^2 \,.
\end{align}
See Refs.~\cite{Contaldi:2003zv,Starobinsky:1992ts,Dvorkin:2009ne,Ballardini:2016hpi} 
for a clear descripition of the models.

\begin{figure}
\centering
\includegraphics[width=.5\textwidth]{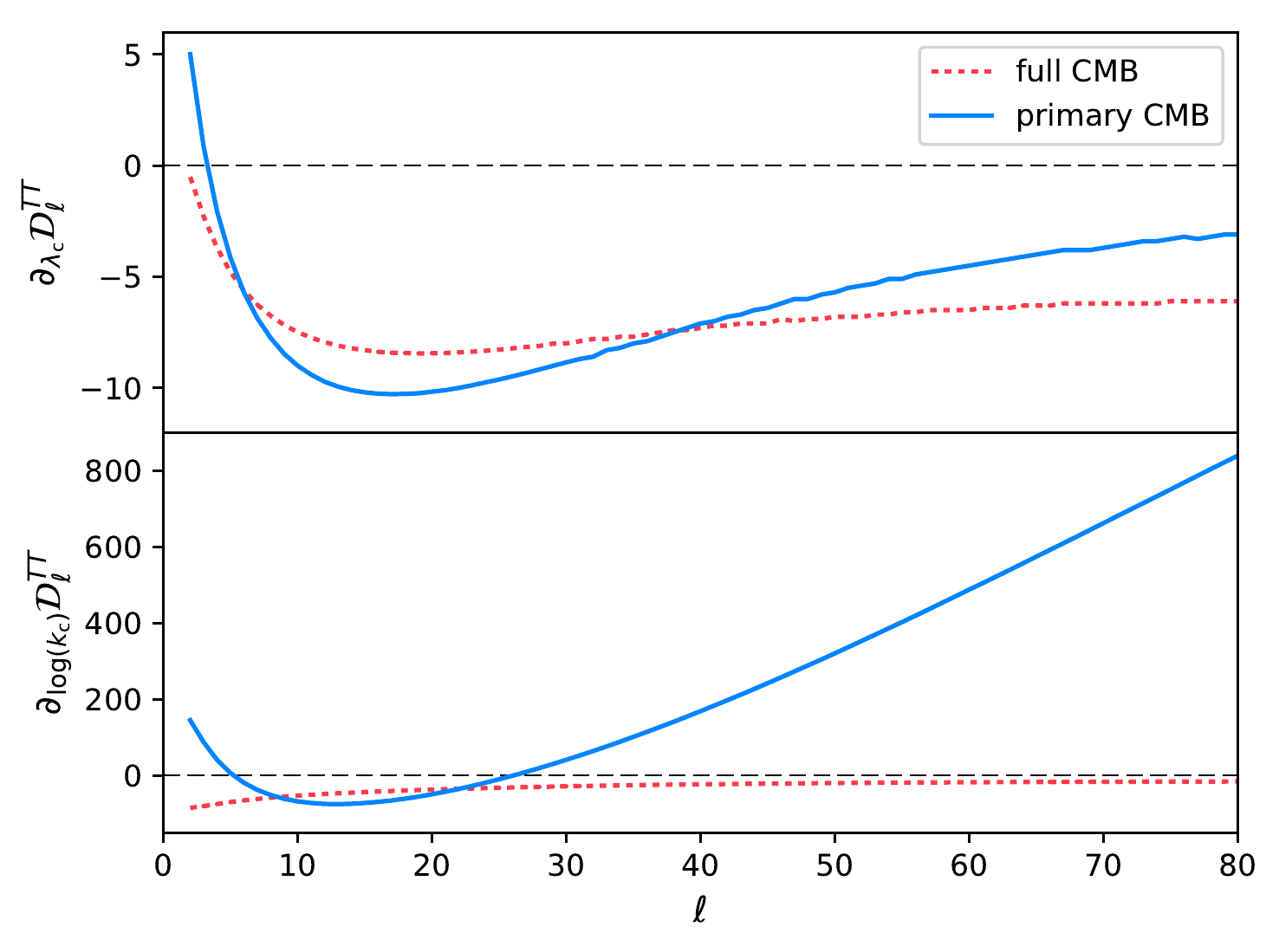}
\includegraphics[width=.5\textwidth]{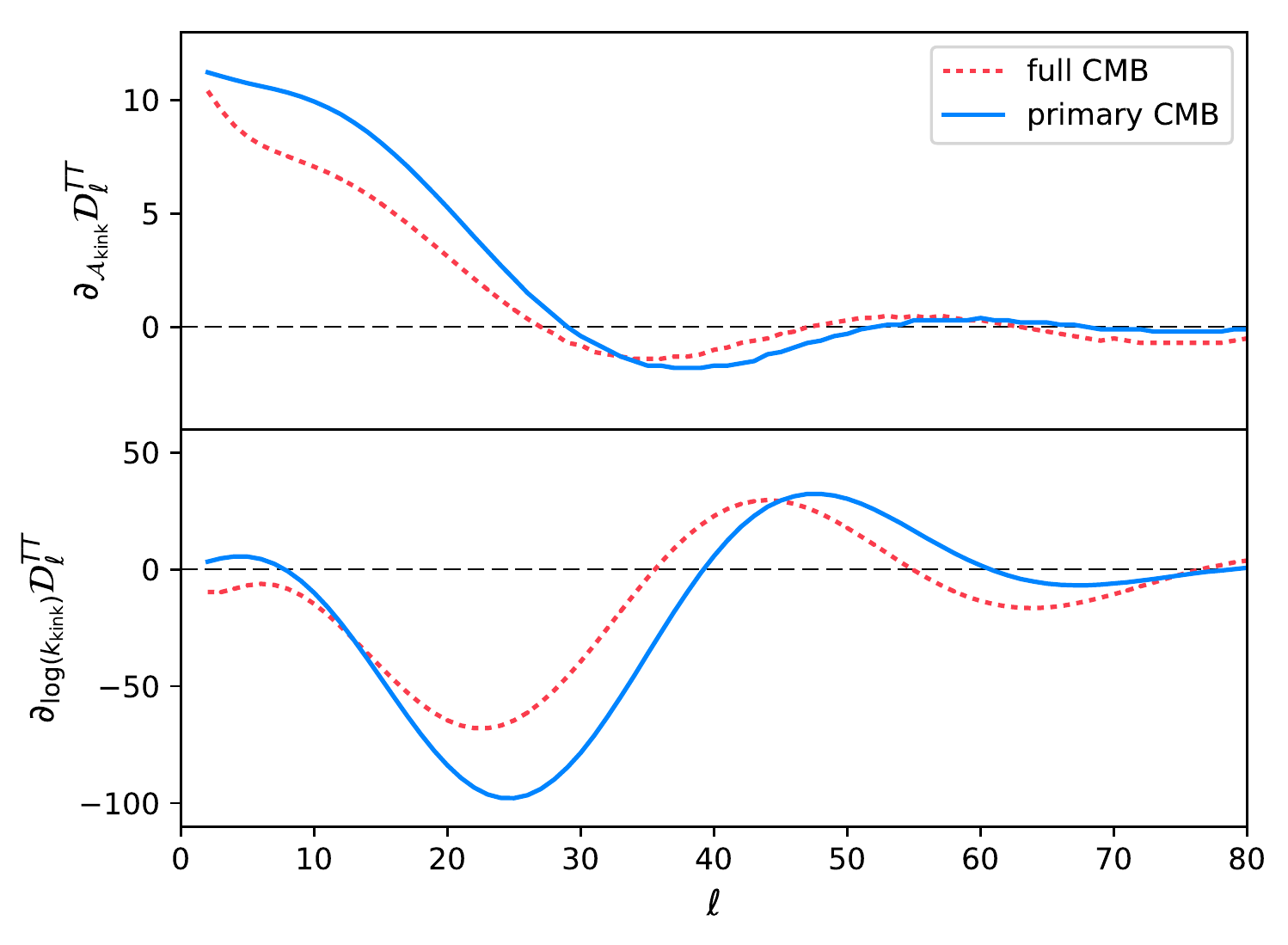}
\includegraphics[width=.5\textwidth]{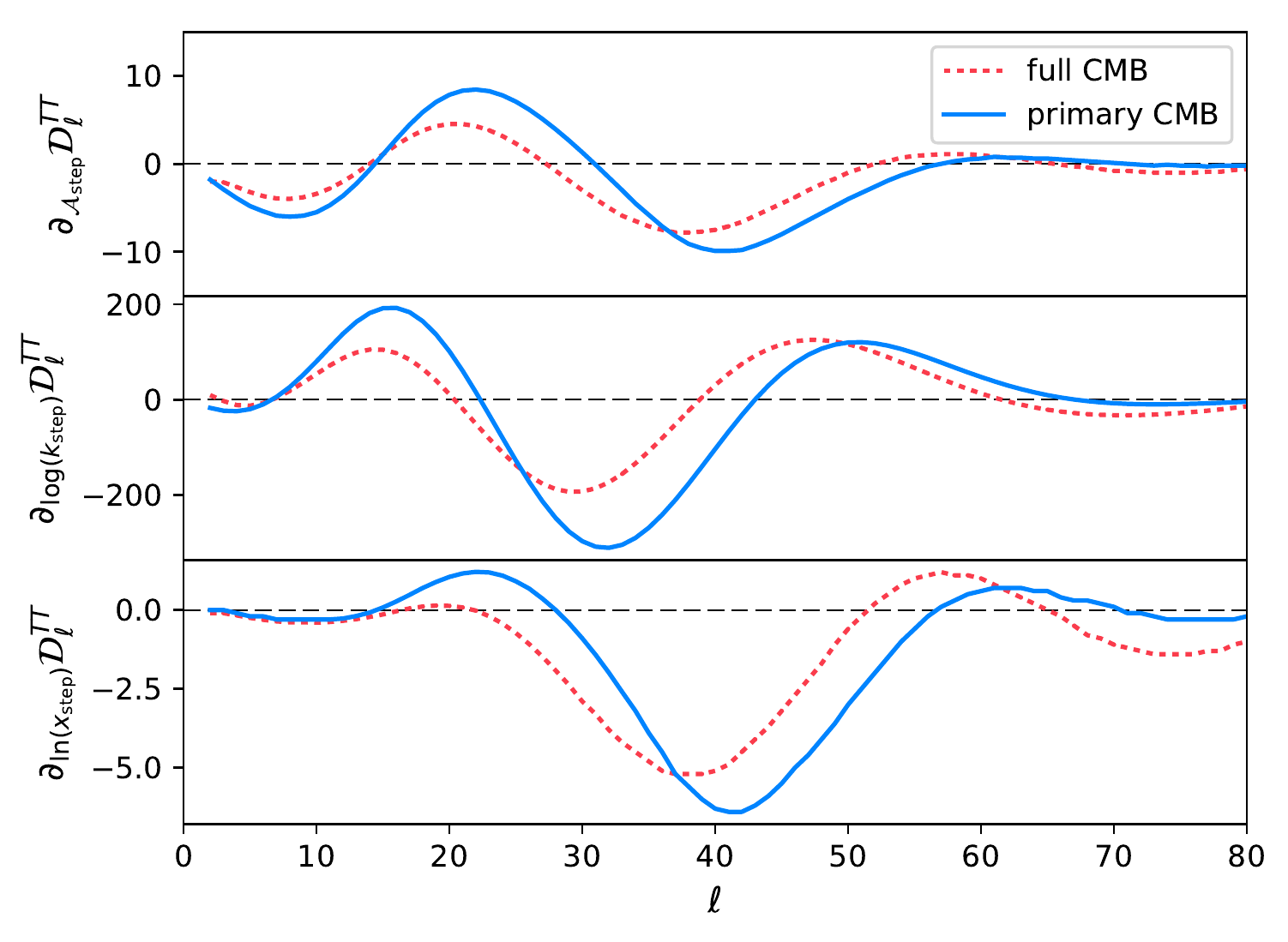}
\caption{Derivatives of the CMB temperature angular power spectrum with respect to the 
features parameters for the {\bf cut-off} (top panels), {\bf kink} (central panels), 
{\bf step} (bottom panels) models. The red dashed lines refer to the derivative of the full 
observed CMB temperature angular power spectrum and the blue solid lines refer to the derivative 
of the primary spectra.}
\label{fig:der1}
\end{figure}

\section{Results}
After the ISW removal, the signal of the CMB temperature anisotropies decreases at the 
large angular scales. On these scales where the instrumental noise is negligible, this 
effect is compensated by an effective lower cosmic-variance since $\sqrt{2/(2\ell+1)}C_\ell$, 
i.e.
\be
\frac{C_\ell^{\rm TT}}{C_\ell^{\rm TT}+{\cal N}_\ell^{\rm T}} \approx 
\frac{C_\ell^{\rm primary}}{C_\ell^{\rm primary}+{\cal N}_\ell^{\rm primary}} \,,
\ee
assuming a negligible ${\cal N}_\ell^{\rm ISW}$. On the other hand, the primary CMB is 
more sensible to the variation of the cosmological parameters connected with the primordial 
power spectrum.
It is possible to see this effect by looking at the derivatives of the CMB temperature 
anisotropies. In Fig.~\ref{fig:der1}, the derivatives of the CMB primary anisotropies respect 
to the features parameters are always larger in amplitude compared to the derivatives of the 
observed CMB temperature anisotropies.

We consider two different configurations of CMB experiment: a reprentative of current CMB 
measurements by considering the $Planck$ 143 GHz channel full mission sensitivity and angular 
resolution as given in \cite{Adam:2015rua} and a CMB cosmic-variance limited experiment, 
both with $f_{\rm sky}=0.7$. Results are collected in Tab.~\ref{tab:results}.

Assuming a perfect reconstruction of the ISW, we found that for the {\bf cut-off} 
model the errors decrease by 6\% on $\lambda_{\rm c}$ and by 16\% on 
$\log_{10}\left(k_{\rm c}\,{\rm Mpc}^{-1}\right)$ for an experiment with $Planck$'s sensitivity. 
For the {\bf kink} model the errors improve by a 17\% on ${\cal A}_{\rm kink}$ and by 10\% on 
$\log_{10}\left(k_{\rm kink}\,{\rm Mpc}^{-1}\right)$. 
The {\bf step} model is the one that benefits more from this method, the errors improve by 
a 27\% on ${\cal A}_{\rm step}$, by 25\% on $\log_{10}\left(k_{\rm step}\,{\rm Mpc}^{-1}\right)$ 
and by 19\% on $\ln\left(x_{\rm step}\right)$.

The case with the subtraction of the 3$\sigma$ and of the $6\sigma$ detected ISW does not lead 
to any improvements for both the {\bf cut-off} and the {\bf kink} models.
Instead, for the {\bf step} model there is still a reduced improvement of 5\% on the amplitude 
and 10\% on the scale parameter, even for these cases with injected noise from the ISW reconstruction.

Feature models which predict departures from the standard power-law primordial power spectrum 
will benefit from having better measurements of large angular scale CMB $E$-mode polarization 
at the cosmic-variance level.
However, the {\bf cut-off} model affects the largest angular scales reproducing a suppression 
of power at $\ell < 30$ in temperature and $\ell < 10$ in the $E$-mode polarization. For this reason, 
the relative improvement does not change when we consider a cosmic-variance limited CMB experiment. 
The instrumental noise on the $E$-mode polarization is small even for $Planck$ on such scales.
In this case the improvement from the subtraction of the ISW signal is very small, $\sim 5\%$, 
even for the case of perfect ISW reconstruction for all the three considered models. 

\section{Conclusion}
In conclusion, this method represent a consistent way to test the origin of features in the 
CMB temperature angular power spectrum. 

We show that the constraints on features parameters after the ISW subtraction are expected 
to be consistent with the ones obtained from full CMB.
Moreover, this approach performs well for the {\bf step} model which fits the deficit in 
power at $\ell \simeq 20-30$, improving the contraints by $5-27\%$ on the amplitude and 
by $10-25\%$ on the scale parameters, even without better measurements of CMB polarization.

Finally, even if the final improvement for realistic cases of ISW subtraction could lead to 
small differences in terms of constraining power on the parameters of these features models, 
the subtraction of the ISW signal could lead to a change in the pattern of the largest scales 
of the CMB temperature anisotropies changing the shape of the features. 
For instance, if an anomaly vanishes after the subtraction of the ISW component to the 
CMB temperature, then a primordial explanation would be eliminated.

\begin{table*}
\vspace{3mm}
\begin{tabular}{|cc|cccc|}
\hline
Model & Parameters & full CMB & $3\sigma$ ISW & $6\sigma$ ISW & primary CMB \\
\hline
\multirow{2}{*}{\bf cut-off} & $\lambda_{\rm c}=0.50$ & $0.218/0.176$ & $0.340/0.234$ & $0.264/0.196$ & $0.204/0.162$ \\ & $\log_{10}\left(k_{\rm c}\,\text{Mpc}^{-1}\right)=-3.47$ & $0.371/0.325$ & $0.564/0.440$ & $0.418/0.357$ & $0.310/0.280$ \\ 
\hline
\multirow{2}{*}{\bf kink} & ${\cal A}_{\rm kink}=0.089$ & $0.0466/0.0334$ & $0.0779/0.0430$ & $0.534/0.0360$ & $0.0387/0.0296$ \\ & $\log_{10}\left(k_{\rm kink}\,\text{Mpc}^{-1}\right)=-3.05$ & $0.0962/0.0530$ & $0.129/0.0564$ & $0.105/0.0549$ & $0.0866/0.0529$ \\ 
\hline
\multirow{3}{*}{\bf step} & ${\cal A}_{\rm step}=0.374$ & $0.257/0.125$ & $0.344/0.130$ & $0.247/0.126$ & $0.187/0.120$ \\ & $\log_{10}\left(k_{\rm step}\,\text{Mpc}^{-1}\right)=-3.1$ & $0.0368/0.0165$ & $0.0418/0.0174$ & $0.0336/0.0168$ & $0.0275/0.0160$ \\ & $\ln\left(x_{\rm step}\right)=0.342$ & $0.362/0.189$ & $0.471/0.199$ & $0.364/0.192$ & $0.293/0.183$ \\ 
\hline
\end{tabular}
\caption{68\% constraints on the features parameters for a $Planck$-like CMB experiment (left) 
and a cosmic-variance limited one (right). Constraints are given for the standard case (full CMB) 
and after ISW subtraction by considering different levels of ISW detection. We report the 
best-fit for the features parameters from $Planck$ 2015 TT + lowP data \cite{Ade:2015lrj}.}
\label{tab:results}
\end{table*}

\section*{Acknowledgments}
We thank Xingang Chen and Roy Maartens for helpful discussions, and Fabio Finelli.
This research was supported by the South African Radio Astronomy Observatory, which 
is a facility of the National Research Foundation, an agency of the 
Department of Science and Technology, and was also supported by the Claude Leon Foundation.
We acknowledge support by the ASI n.I/023/12/0 "Attivit\`a relative alla fase B2/C per 
la missione Euclid".
We would like to thank Johns Hopkins University and Harvard University for the hospitality.



\begin{thebibliography}{99}

\bibitem{Akrami:2018vks}
  Y.~Akrami {\it et al.} [Planck Collaboration],
  arXiv:1807.06205 [astro-ph.CO].



\bibitem{Peiris:2003ff}
  H.~V.~Peiris {\it et al.} [WMAP Collaboration],
  Astrophys.\ J.\ Suppl.\  {\bf 148} (2003) 213
  doi:10.1086/377228
  [astro-ph/0302225].



\bibitem{Planck:2013jfk}
  P.~A.~R.~Ade {\it et al.} [Planck Collaboration],
  Astron.\ Astrophys.\  {\bf 571} (2014) A22
  doi:10.1051/0004-6361/201321569
  [arXiv:1303.5082 [astro-ph.CO]].



\bibitem{Ade:2015lrj}
  P.~A.~R.~Ade {\it et al.} [Planck Collaboration],
  Astron.\ Astrophys.\  {\bf 594} (2016) A20
  doi:10.1051/0004-6361/201525898
  [arXiv:1502.02114 [astro-ph.CO]].



\bibitem{Akrami:2018odb}
  Y.~Akrami {\it et al.} [Planck Collaboration],
  arXiv:1807.06211 [astro-ph.CO].



\bibitem{Covi:2006ci}
  L.~Covi, J.~Hamann, A.~Melchiorri, A.~Slosar and I.~Sorbera,
  Phys.\ Rev.\ D {\bf 74} (2006) 083509
  doi:10.1103/PhysRevD.74.083509
  [astro-ph/0606452].



\bibitem{Benetti:2013cja}
  M.~Benetti,
  Phys.\ Rev.\ D {\bf 88} (2013) 087302
  doi:10.1103/PhysRevD.88.087302
  [arXiv:1308.6406 [astro-ph.CO]].



\bibitem{Miranda:2013wxa}
  V.~Miranda and W.~Hu,
  Phys.\ Rev.\ D {\bf 89} (2014) no.8,  083529
  doi:10.1103/PhysRevD.89.083529
  [arXiv:1312.0946 [astro-ph.CO]].



\bibitem{Easther:2013kla}
  R.~Easther and R.~Flauger,
  JCAP {\bf 1402} (2014) 037
  doi:10.1088/1475-7516/2014/02/037
  [arXiv:1308.3736 [astro-ph.CO]].



\bibitem{Chen:2014joa}
  X.~Chen and M.~H.~Namjoo,
  Phys.\ Lett.\ B {\bf 739} (2014) 285
  doi:10.1016/j.physletb.2014.11.002
  [arXiv:1404.1536 [astro-ph.CO]].



\bibitem{Achucarro:2014msa}
  A.~Achucarro, V.~Atal, B.~Hu, P.~Ortiz and J.~Torrado,
  Phys.\ Rev.\ D {\bf 90} (2014) no.2,  023511
  doi:10.1103/PhysRevD.90.023511
  [arXiv:1404.7522 [astro-ph.CO]].



\bibitem{Hazra:2014goa}
  D.~K.~Hazra, A.~Shafieloo, G.~F.~Smoot and A.~A.~Starobinsky,
  JCAP {\bf 1408} (2014) 048
  doi:10.1088/1475-7516/2014/08/048
  [arXiv:1405.2012 [astro-ph.CO]].



\bibitem{Hazra:2014jwa}
  D.~K.~Hazra, A.~Shafieloo and T.~Souradeep,
  JCAP {\bf 1411} (2014) no.11,  011
  doi:10.1088/1475-7516/2014/11/011
  [arXiv:1406.4827 [astro-ph.CO]].



\bibitem{Hu:2014hra}
  B.~Hu and J.~Torrado,
  Phys.\ Rev.\ D {\bf 91} (2015) no.6,  064039
  doi:10.1103/PhysRevD.91.064039
  [arXiv:1410.4804 [astro-ph.CO]].



\bibitem{Gruppuso:2015zia}
  A.~Gruppuso and A.~Sagnotti,
  Int.\ J.\ Mod.\ Phys.\ D {\bf 24} (2015) no.12,  1544008
  doi:10.1142/S0218271815440083
  [arXiv:1506.08093 [astro-ph.CO]].



\bibitem{Gruppuso:2015xqa}
  A.~Gruppuso, N.~Kitazawa, N.~Mandolesi, P.~Natoli and A.~Sagnotti,
  Phys.\ Dark Univ.\  {\bf 11} (2016) 68
  doi:10.1016/j.dark.2015.12.001
  [arXiv:1508.00411 [astro-ph.CO]].



\bibitem{Hazra:2016fkm}
  D.~K.~Hazra, A.~Shafieloo, G.~F.~Smoot and A.~A.~Starobinsky,
  JCAP {\bf 1609} (2016) no.09,  009
  doi:10.1088/1475-7516/2016/09/009
  [arXiv:1605.02106 [astro-ph.CO]].



\bibitem{Torrado:2016sls}
  J.~Torrado, B.~Hu and A.~Achucarro,
  Phys.\ Rev.\ D {\bf 96} (2017) no.8,  083515
  doi:10.1103/PhysRevD.96.083515
  [arXiv:1611.10350 [astro-ph.CO]].



\bibitem{Obied:2018qdr}
  G.~Obied, C.~Dvorkin, C.~Heinrich, W.~Hu and V.~Miranda,
  Phys.\ Rev.\ D {\bf 98} (2018) no.4,  043518
  doi:10.1103/PhysRevD.98.043518
  [arXiv:1803.01858 [astro-ph.CO]].



\bibitem{Mortonson:2009qv}
  M.~J.~Mortonson, C.~Dvorkin, H.~V.~Peiris and W.~Hu,
  Phys.\ Rev.\ D {\bf 79} (2009) 103519
  doi:10.1103/PhysRevD.79.103519
  [arXiv:0903.4920 [astro-ph.CO]].



\bibitem{Chluba:2015bqa}
  J.~Chluba, J.~Hamann and S.~P.~Patil,
  Int.\ J.\ Mod.\ Phys.\ D {\bf 24} (2015) no.10,  1530023
  doi:10.1142/S0218271815300232
  [arXiv:1505.01834 [astro-ph.CO]].



\bibitem{Finelli:2016cyd}
  F.~Finelli {\it et al.} [CORE Collaboration],
  JCAP {\bf 1804} (2018) 016
  doi:10.1088/1475-7516/2018/04/016
  [arXiv:1612.08270 [astro-ph.CO]].



\bibitem{Hazra:2017joc}
  D.~K.~Hazra, D.~Paoletti, M.~Ballardini, F.~Finelli, A.~Shafieloo, G.~F.~Smoot and A.~A.~Starobinsky,
  JCAP {\bf 1802} (2018) no.02,  017
  doi:10.1088/1475-7516/2018/02/017
  [arXiv:1710.01205 [astro-ph.CO]].



\bibitem{Zhan:2005rz}
  H.~Zhan, L.~Knox, A.~Tyson and V.~Margoniner,
  Astrophys.\ J.\  {\bf 640} (2006) 8
  doi:10.1086/500077
  [astro-ph/0508119].



\bibitem{Huang:2012mr}
  Z.~Huang, L.~Verde and F.~Vernizzi,
  JCAP {\bf 1204} (2012) 005
  doi:10.1088/1475-7516/2012/04/005
  [arXiv:1201.5955 [astro-ph.CO]].



\bibitem{Chen:2016vvw}
  X.~Chen, C.~Dvorkin, Z.~Huang, M.~H.~Namjoo and L.~Verde,
  JCAP {\bf 1611} (2016) no.11,  014
  doi:10.1088/1475-7516/2016/11/014
  [arXiv:1605.09365 [astro-ph.CO]].



\bibitem{Chen:2016zuu}
  X.~Chen, P.~D.~Meerburg and M.~Münchmeyer,
  JCAP {\bf 1609} (2016) no.09,  023
  doi:10.1088/1475-7516/2016/09/023
  [arXiv:1605.09364 [astro-ph.CO]].



\bibitem{Ballardini:2016hpi}
  M.~Ballardini, F.~Finelli, C.~Fedeli and L.~Moscardini,
  JCAP {\bf 1610} (2016) 041
   Erratum: [JCAP {\bf 1804} (2018) no.04,  E01]
  doi:10.1088/1475-7516/2018/04/E01, 10.1088/1475-7516/2016/10/041
  [arXiv:1606.03747 [astro-ph.CO]].



\bibitem{Xu:2016kwz}
  Y.~Xu, J.~Hamann and X.~Chen,
  Phys.\ Rev.\ D {\bf 94} (2016) no.12,  123518
  doi:10.1103/PhysRevD.94.123518
  [arXiv:1607.00817 [astro-ph.CO]].



\bibitem{Fard:2017oex}
  M.~A.~Fard and S.~Baghram,
  JCAP {\bf 1801} (2018) no.01,  051
  doi:10.1088/1475-7516/2018/01/051
  [arXiv:1709.05323 [astro-ph.CO]].



\bibitem{Palma:2017wxu}
  G.~A.~Palma, D.~Sapone and S.~Sypsas,
  JCAP {\bf 1806} (2018) no.06,  004
  doi:10.1088/1475-7516/2018/06/004
  [arXiv:1710.02570 [astro-ph.CO]].



\bibitem{LHuillier:2017lgm}
  B.~L'Huillier, A.~Shafieloo, D.~K.~Hazra, G.~F.~Smoot and A.~A.~Starobinsky,
  Mon.\ Not.\ Roy.\ Astron.\ Soc.\  {\bf 477} (2018) no.2,  2503
  doi:10.1093/mnras/sty745
  [arXiv:1710.10987 [astro-ph.CO]].



\bibitem{Ballardini:2017qwq}
  M.~Ballardini, F.~Finelli, R.~Maartens and L.~Moscardini,
  JCAP {\bf 1804} (2018) no.04,  044
  doi:10.1088/1475-7516/2018/04/044
  [arXiv:1712.07425 [astro-ph.CO]].



\bibitem{Fergusson:2014hya}
  J.~R.~Fergusson, H.~F.~Gruetjen, E.~P.~S.~Shellard and M.~Liguori,
  Phys.\ Rev.\ D {\bf 91} (2015) no.2,  023502
  doi:10.1103/PhysRevD.91.023502
  [arXiv:1410.5114 [astro-ph.CO]].



\bibitem{Fergusson:2014tza}
  J.~R.~Fergusson, H.~F.~Gruetjen, E.~P.~S.~Shellard and B.~Wallisch,
  Phys.\ Rev.\ D {\bf 91} (2015) no.12,  123506
  doi:10.1103/PhysRevD.91.123506
  [arXiv:1412.6152 [astro-ph.CO]].



\bibitem{Ade:2015ava}
  P.~A.~R.~Ade {\it et al.} [Planck Collaboration],
  Astron.\ Astrophys.\  {\bf 594} (2016) A17
  doi:10.1051/0004-6361/201525836
  [arXiv:1502.01592 [astro-ph.CO]].



\bibitem{Meerburg:2015owa}
  P.~D.~Meerburg, M.~Münchmeyer and B.~Wandelt,
  Phys.\ Rev.\ D {\bf 93} (2016) no.4,  043536
  doi:10.1103/PhysRevD.93.043536
  [arXiv:1510.01756 [astro-ph.CO]].



\bibitem{MoradinezhadDizgah:2018ssw}
  A.~Moradinezhad Dizgah, H.~Lee, J.~B.~Muñoz and C.~Dvorkin,
  JCAP {\bf 1805} (2018) no.05,  013
  doi:10.1088/1475-7516/2018/05/013
  [arXiv:1801.07265 [astro-ph.CO]].



\bibitem{Karagiannis:2018jdt}
  D.~Karagiannis, A.~Lazanu, M.~Liguori, A.~Raccanelli, N.~Bartolo and L.~Verde,
  Mon.\ Not.\ Roy.\ Astron.\ Soc.\  {\bf 478} (2018) no.1,  1341
  doi:10.1093/mnras/sty1029
  [arXiv:1801.09280 [astro-ph.CO]].



\bibitem{Efstathiou:2009di}
  G.~Efstathiou, Y.~Z.~Ma and D.~Hanson,
  Mon.\ Not.\ Roy.\ Astron.\ Soc.\  {\bf 407} (2010) 2530
  doi:10.1111/j.1365-2966.2010.17081.x
  [arXiv:0911.5399 [astro-ph.CO]].



\bibitem{Francis:2009pt}
  C.~L.~Francis and J.~A.~Peacock,
  Mon.\ Not.\ Roy.\ Astron.\ Soc.\  {\bf 406} (2010) 14
  doi:10.1111/j.1365-2966.2010.16866.x
  [arXiv:0909.2495 [astro-ph.CO]].



\bibitem{Muir:2016veb}
  J.~Muir and D.~Huterer,
  Phys.\ Rev.\ D {\bf 94} (2016) no.4,  043503
  doi:10.1103/PhysRevD.94.043503
  [arXiv:1603.06586 [astro-ph.CO]].



\bibitem{Copi:2016hhq}
  C.~J.~Copi, M.~O'Dwyer and G.~D.~Starkman,
  Mon.\ Not.\ Roy.\ Astron.\ Soc.\  {\bf 463} (2016) no.3,  3305
  doi:10.1093/mnras/stw2163
  [arXiv:1605.09732 [astro-ph.CO]].



\bibitem{Kofman:1985fp}
  L.~Kofman and A.~A.~Starobinsky,
  Sov.\ Astron.\ Lett.\  {\bf 11} (1985) 271
   [Pisma Astron.\ Zh.\  {\bf 11} (1985) 643].



\bibitem{Crittenden:1995ak}
  R.~G.~Crittenden and N.~Turok,
  Phys.\ Rev.\ Lett.\  {\bf 76} (1996) 575
  doi:10.1103/PhysRevLett.76.575
  [astro-ph/9510072].



\bibitem{Manzotti:2014kta}
  A.~Manzotti and S.~Dodelson,
  Phys.\ Rev.\ D {\bf 90} (2014) no.12,  123009
  doi:10.1103/PhysRevD.90.123009
  [arXiv:1407.5623 [astro-ph.CO]].



\bibitem{Taburet:2010hb}
  N.~Taburet, C.~Hernandez-Monteagudo, N.~Aghanim, M.~Douspis and R.~A.~Sunyaev,
  Mon.\ Not.\ Roy.\ Astron.\ Soc.\  {\bf 418} (2011) 2207
  doi:10.1111/j.1365-2966.2011.19474.x
  [arXiv:1012.5036 [astro-ph.CO]].



\bibitem{Pourtsidou:2016dzn}
  A.~Pourtsidou, D.~Bacon and R.~Crittenden,
  Mon.\ Not.\ Roy.\ Astron.\ Soc.\  {\bf 470} (2017) no.4,  4251
  doi:10.1093/mnras/stx1479
  [arXiv:1610.04189 [astro-ph.CO]].



\bibitem{Ballardini:2017xnt}
  M.~Ballardini, D.~Paoletti, F.~Finelli, L.~Moscardini, B.~Sartoris and L.~Valenziano,
  arXiv:1712.02380 [astro-ph.CO].



\bibitem{Cooray:2001ab}
  A.~Cooray,
  Phys.\ Rev.\ D {\bf 65} (2002) 103510
  doi:10.1103/PhysRevD.65.103510
  [astro-ph/0112408].



\bibitem{Afshordi:2004kz}
  N.~Afshordi,
  Phys.\ Rev.\ D {\bf 70} (2004) 083536
  doi:10.1103/PhysRevD.70.083536
  [astro-ph/0401166].



\bibitem{Ade:2015dva}
  P.~A.~R.~Ade {\it et al.} [Planck Collaboration],
  Astron.\ Astrophys.\  {\bf 594} (2016) A21
  doi:10.1051/0004-6361/201525831
  [arXiv:1502.01595 [astro-ph.CO]].



\bibitem{Raccanelli:2015lca}
  A.~Raccanelli, E.~Kovetz, L.~Dai and M.~Kamionkowski,
  Phys.\ Rev.\ D {\bf 93} (2016) no.8,  083512
  doi:10.1103/PhysRevD.93.083512
  [arXiv:1502.03107 [astro-ph.CO]].



\bibitem{Knox:1995dq}
  L.~Knox,
  Phys.\ Rev.\ D {\bf 52} (1995) 4307
  doi:10.1103/PhysRevD.52.4307
  [astro-ph/9504054].



\bibitem{Jungman:1995bz}
  G.~Jungman, M.~Kamionkowski, A.~Kosowsky and D.~N.~Spergel,
  Phys.\ Rev.\ D {\bf 54} (1996) 1332
  doi:10.1103/PhysRevD.54.1332
  [astro-ph/9512139].



\bibitem{Seljak:1996ti}
  U.~Seljak,
  Astrophys.\ J.\  {\bf 482} (1997) 6
  doi:10.1086/304123
  [astro-ph/9608131].



\bibitem{Zaldarriaga:1996xe}
  M.~Zaldarriaga and U.~Seljak,
  Phys.\ Rev.\ D {\bf 55} (1997) 1830
  doi:10.1103/PhysRevD.55.1830
  [astro-ph/9609170].



\bibitem{Kamionkowski:1996ks}
  M.~Kamionkowski, A.~Kosowsky and A.~Stebbins,
  Phys.\ Rev.\ D {\bf 55} (1997) 7368
  doi:10.1103/PhysRevD.55.7368
  [astro-ph/9611125].



\bibitem{Contaldi:2003zv}
  C.~R.~Contaldi, M.~Peloso, L.~Kofman and A.~D.~Linde,
  JCAP {\bf 0307} (2003) 002
  doi:10.1088/1475-7516/2003/07/002
  [astro-ph/0303636].



\bibitem{Starobinsky:1992ts}
  A.~A.~Starobinsky,
  JETP Lett.\  {\bf 55} (1992) 489
   [Pisma Zh.\ Eksp.\ Teor.\ Fiz.\  {\bf 55} (1992) 477].



\bibitem{Dvorkin:2009ne}
  C.~Dvorkin and W.~Hu,
  Phys.\ Rev.\ D {\bf 81} (2010) 023518
  doi:10.1103/PhysRevD.81.023518
  [arXiv:0910.2237 [astro-ph.CO]].



\bibitem{Adam:2015rua}
  R.~Adam {\it et al.} [Planck Collaboration],
  Astron.\ Astrophys.\  {\bf 594} (2016) A1
  doi:10.1051/0004-6361/201527101
  [arXiv:1502.01582 [astro-ph.CO]].





\end{thebibliography}
\end{document}